\documentclass[10pt,a4paper,twocolumn]{article}
\usepackage{graphicx} 
\usepackage{amstext} 
\usepackage[top=25mm,bottom=15mm,left=20mm,right=20mm,columnsep=10mm]{geometry} 
\usepackage{color} 
\definecolor{myblu}{rgb}{0.1,0.1,0.5}
\usepackage{hyperref} 
\hypersetup{colorlinks=true,urlcolor=blue,linkcolor=black,citecolor=black} 
\usepackage{sectsty,textcase} 
\sectionfont{\large\MakeTextUppercase} 
\usepackage{secdot} 

\usepackage{float}
\usepackage{graphicx}
\usepackage{subfigmat}

\usepackage{amsmath, amssymb, graphics, setspace}
\usepackage{amsfonts}

\begin{document}
\title{\vspace{-18mm}
\begin{minipage}{\linewidth}
\hspace{5mm}\raisebox{-50pt}{\includegraphics[width=.23\textwidth]{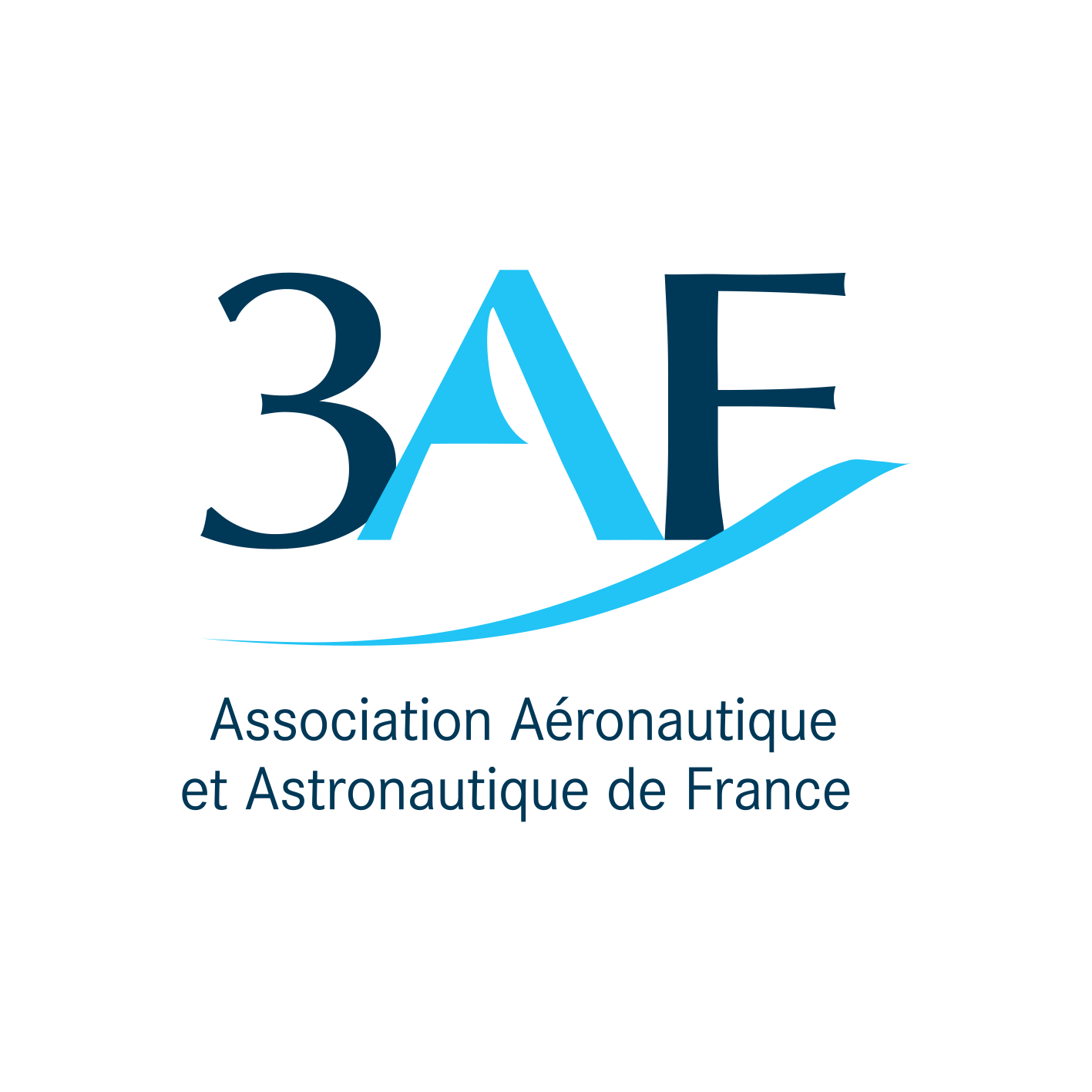}}\hspace{4mm}
\textcolor{myblu}{\textbf{\textit{\normalsize\begin{tabular}{l}
8$^\text{th}$ 3AF Space Propulsion Conference\\ 9 --- 13 May 2022, Estoril -- Portugal
\end{tabular}}}}
\hspace{16mm}\textbf{\normalsize SP2022-452}
\end{minipage}\\\vspace{10mm}
\textbf{\large European Shock-Tube for High-Enthalpy Research Combustion Driver Qualification}}
\author{\textbf{\normalsize Ricardo Grosso Ferreira$^\text{(1)}$ and Mário Lino da Silva$^\text{(2)}$}
\\{\normalsize\itshape
$^\text{(1)}$IPFN, Av. Rovisco Pais nº1 1049-001 Lisboa, Portugal, ricardojoaogmferreira@tecnico.ulisboa.pt}
\\{\normalsize\itshape
$^\text{(2)}$IPFN, Av. Rovisco Pais nº1 1049-001 Lisboa, Portugal, mlinodasilva@tecnico.ulisboa.pt}}
\date{}

\maketitle
\begin{abstract}

The ESTHER shock tube is a new state-of-the-art facility at Instituto Superior Técnico designed to support future ESA planetary exploration missions. Its driver is a high-pressure combustion chamber using a mixture of He:H$_2$:O$_2$ ignited by a high-power Nd:YAG laser. Both hydrogen as an energy vector and laser ignition are promising techniques with applications in high-pressure combustion. The influence of gas mixture and laser parameters, namely the air:fuel ratio, filling pressure, inert gas dilution and ignition mode, on the combustion and thus shock tube performance were extensively studied. A second, low-velocity driver mixture with nitrogen in place of helium as a dilutant was also studied and experimental shots are done. 

Our results show that the filling pressure and helium dilution are the most dominant parameters in both peak pressure, acoustic oscillation and combustion velocity. The gas mixture peak pressure and acoustic wave amplitude increase with the increased filling pressure. Yet, the increased filling pressure lowers the combustion velocity. The helium in the mixture had a dilution effect, with it lowering the overall effectiveness of combustion. Having higher dilution factors lowers the combustion compression ratio, acoustic waves amplitude and flame velocity. The air:fuel equivalence ratio influence was expected with faster flame and peak pressures at the stoichiometric region. Nitrogen diluted shots have drastically lower compression ratios and flame velocity when compared to the helium ones, besides it, the acoustic perturbation was stronger. ``tulip" flames and deflagration to detonation transitions phenomena were identified in some of the experiments.


%
%
\end{abstract}

\section{Introduction}

\begin{figure*}[ht]
\centering
\includegraphics[width=0.99\textwidth]{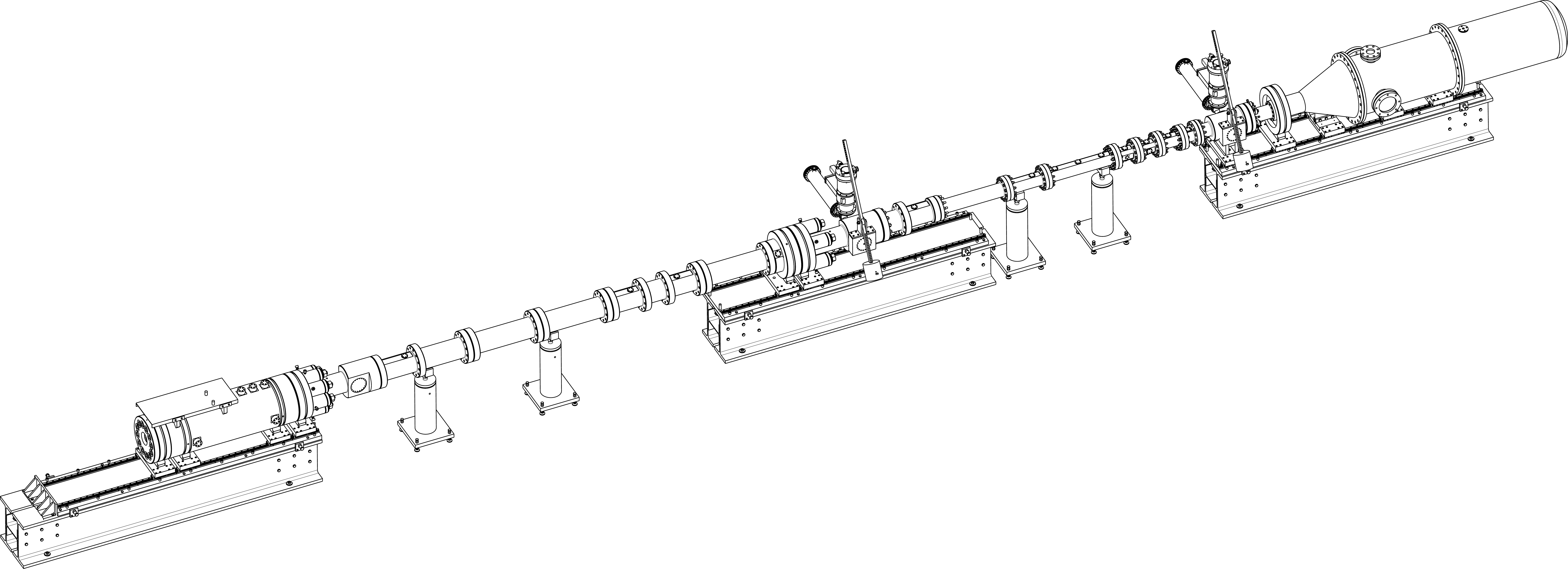}
\caption{ESTHER CAD overview}
\label{Fig: ESTHER_overview}
\end{figure*}

The European Shock-Tube for High-Enthalpy Research (ESTHER) is a new state-of-the-art facility at Instituto Superior Técnico design to support the next European Space Agency (ESA) planetary exploration missions \cite{Lino-Chikhaoui-ESTHER-2012,Diana-Luis-MSc-Thesis}. A shock-tube is comprised of a high-pressure driver, and a low-pressure driven sections separated by a breakable wall, a diaphgram. The driver pressure increases until it ruptures the diaphgram creating a high-speed shock-wave. This may be achieved using multiple techniques \cite{Reynier2016}, namely a high-pressure combustion driver. Hydrogen arises as a natural solution for a high-pressure combustion shock-tube driver due to being lighter and not producing soot when compared to hydrocarbons. Another advantage of hydrogen in relation to hydrocarbons is that its combustion temperature and peak pressure are higher than most hydrocarbons, furthering improving the driver performance. Due to the high filling pressures (10-100 bar), laser ignition provides significant advantages over electric spark plugs. The most significant one is that increasing filling pressure increases the minimum voltage in the spark plugs, yet, decreases the minimum pulse energy as observed in \cite{Srivastava-2009,Phuoc-White,Phuoc-2006,Kopecek-2003,Kopecek-Reider-2003}. To accomplish good repeatability the combustion should be stable and with subsonic flame propagation mode (deflagration) instead of supersonic (detonation). Strong acoustic oscillations are also undesirable, thus a small scale model of the chamber was first tested in \cite{Mario-Lino-ESTHER-Quali-2020}, yielding an operational mixture of He:H$_2$:O$_2$ 8:2:1.2. A qualification campaign went underway to ensure the driver predictability, repeatability and reliability. A mapping of the limits of deflagration and detonation, acoustic oscillations, and peak pressure as function of the initial conditions, pressure $p_0$, dilution $X_{He}$ and air:fuel ratio $\lambda$, and, of the laser focusing lens. Fig. \ref{Fig: ESTHER_overview} depicts ESTHER CAD overview, with combustion chamber driver on the left and the driven section on the right hand side.

This manuscript is divided into the following parts: Section II covers the state-of-the-art in high-pressure hydrogen combustion; Section III describes the experimental setup; Section IV presents and discusses the results; and Section V highlights the main conclusions and next steps of this work.

\section{State-of-the-Art}

The air:fuel equivalence ratio $\lambda$\footnote{$\phi = \lambda^{-1}$, $\phi$ is the fuel:air equivalence ratio} is defined by the ratio of oxidizer to fuel quantities, normalized by the stoichiometric coefficients. In our experiment, the air:fuel ratio is defined as 

\begin{equation}
\lambda = \left. \frac{n_{O2}}{n_{H2}}  \right/ \left( \frac{n_{O2}}{n_{H2}} \right)_{stoichiometric} \quad,
\label{Eq: Air:Fuel}    
\end{equation}

where $n_{O2}$ and $n_{H2}$ are the number of moles of oxygen and hydrogen respectively. A "rich" mixture is defined by having excess fuel ($\lambda < 1$), and a "lean" mixture by having excess air/oxidizer ($\lambda > 1$) in relation to the stoichiometric conditions ($\lambda = 1$). Maximum peak pressure is typically found at stoichiometric conditions $\lambda \sim 1$.  Reference \cite{Tang-2009} measured the compression ratio and velocity of hydrogen-air-nitrogen mixtures, reporting that peak compression ratio is found at $\lambda=1$ for filling pressures of 1.5 bar. Nonetheless, maximum flame velocity occurs in rich mixtures at $\lambda=0.71$. Hydrogen low molecular weight and high thermal diffusivity explain why hydrogen-rich mixtures will burn faster than stoichiometric ones \cite{Law_Chung_K_Combustion_Physics}.  References \cite{Tse-2001, Burke-2007} observed similar results.

Dilution ratio $X_{He/N2}$ by an inert/non-reactive gas can be expressed as the ratio of non-reactive gas moles to the total gas. In our experiments, these gases can be either helium or nitrogen, thus the Eq.~\ref{Eq: Dilution} follows, 

\begin{equation}
X_{He/N2} = \frac{n_{He/N2}}{n_{total}} \quad .
\label{Eq: Dilution}    
\end{equation}


Compression ratio and flame temperature are not significantly effected by changes in gas filling pressure \cite{Tang-2009}. A small increase in compression ratio may be observed and explained by faster flames, which reduce the wall heat losses, thus ensuring a higher peak pressure/temperature. In combustible mixtures with low to zero dilution by an ``inert"/non-reacting gas faster flames are achieved at increased pressures \cite{Prasad-2017, Burke-2007, Burke-2010, Kuznetsov-2011, Kuznetsov-2012, Tse-2001, Santner-2013}. Nonetheless, this trend is reversed in mixtures with high dilution ($>50\%$) factors. Simulation on hydrogen-oxygen-steam mixtures \cite{Kuznetsov-2011} shows that laminar flame velocity has a non-monotonic dependence with the filling pressure. Similar results are found in \cite{Burke-2010}. Mixtures with dilution above 40\% show a laminar flame velocity nearly inversely proportional to the filling pressure.  In both works the authors justify this behaviour by a change in the kinetic scheme. 

The work of \cite{Schrder2004} shows a non-monotonous pressure dependence of detonation limits for high-pressure hydrogen-oxygen and hydrogen-air mixtures. As stated before, pulsed lasers are effective in igniting high-pressure combustible mixtures, a review of its advantages can be found in \cite{Ronney-1994}. Multiple works on laser ignition of high-pressure mixtures have been performed \cite{Weinrotter-2004, Morsy2003, Qin2000, Srivastava-2009, Dharamshi-2014} for hydrogen-air,  \cite{Spiglanin-et-al-1995} for hydrogen-oxygen and  \cite{Prasad-2017, Kopecek-2003, Phuoc-Single-vs-Multipoint-Laser, Morsy2001, Kopecek2005, Ma-1998, Phuoc-White, Morsy1999, Beduneau2003} methane-air.

Detonation may be initiated directly at ignition or trough a deflagration that transits to a detonation. The detonation region presents very high temperatures and pressures, which leads to exponentially high reaction rates, since these are proportional to the exponential of temperature. A detonation starts by a deposition of a great amount of energy in a short time, like with a focused laser.  On the deflagration-to-detonation transition (DDT), the flame speed is key to the transition. A ``tulip" flame is usually present in DDT, and when the flame front inversion occurs, detonation may start in a preheat ahead of the flame front. Reference \cite{Liberman-2010} illustrates this phenomena in shadow photography, as the pressures waves coalesce ahead of the flame front, creating a region of extremely high pressure. They concluded that the positive feedback loop of pressure and temperature feed the detonation fast reactions, as temperature/pressure increase so the velocity of chemical reactions, which in turn releases more heat to increase the temperature.

A detonation is only self-sustained if the transverse dimension of the tube is larger than the detonation cell, otherwise the detonation will decay to a deflagration. Reference \cite{Lee-1984} used shadowgraph to image detonation cells in hydrogen-oxygen mixtures, evidencing a fish scale pattern. 
Mixtures with smaller cells lead to more sensitive and unstable mixtures, which detonate more easily. The influence of $\lambda$ on the detonation cell size was investigated by \cite{Lee-1984} for several gas mixtures. Detonation cells have a minimum for $\lambda \sim 1$, meaning stoichiometric mixture are easier to detonate and sustain a detonation. Increasing filling pressure decreases the size of detonations cells \cite{Knystautas-1982,Bull-1982}.

In a DDT an initial laminar subsonic flame accelerates and at some point transits to a detonation. This phenomena is a stochastic one, thus the transition point and time is not always the same for the same initial conditions \cite{Liberman-2010}. Experiments on hydrogen-oxygen DDT in tubes were done  in \cite{Oppenheim-1966, Meyer-1970, Meyer-1971}. 
Currently the DDT phenomena happens in the following manner: 1) initial exponential flame acceleration and shockwave creation; 2) flame creates compression waves ahead of its front which grows the shock; 3) A heated zone of unreacted gas right in front of the flame is created, a chain of reactions starts as the flame front reaches it; A positive feedback loop is formed by the rapid heat release of the chemical reaction rates in the high pressure/temperature zone, transit to a detonation. Schlieren photography of stoichiometric hydrogen-oxygen observing the 3 DDT stages previously presented can be found in \cite{Liberman-2010}. A ``tulip" flame was formed during each DDT phenomena, but not all ``tulip" flames transit to a detonation.

A ``tulip" flame is a phenomena where a premixed flame in a tube suddenly changes its flame front from finger-like to a tulip-like shape. In it, the flame front center region slows, thus being ``overtaken" by the outer edges near the tube wall. This was photographed in \cite{Ellis1928}. The ``tulip" formation can be attributed to different causes, such as flow viscosity \cite{Starke-1986, Marra-1996}, flow-pressure wave interaction \cite{Metzener-2001}, hydrodynamic instability \cite{Gonzalez-1992, Dold-1995}, vortices or Taylor instability \cite{MATALON1997, Metzener-2001}. The propagation speed suddenly decreases when the flame front transits to ``tulip"-flame, faster flames create a more pronounced ``tulip", due to the larger acceleration/deceleration, \cite{Xiao-2014}. The flame front inversion may be observed in the signal pressure as the point where a second slope \cite{DunnRankin-1998}.

\section{Experimental Setup}

\begin{figure*}[htb]
\centering
\includegraphics[width=0.99\textwidth]{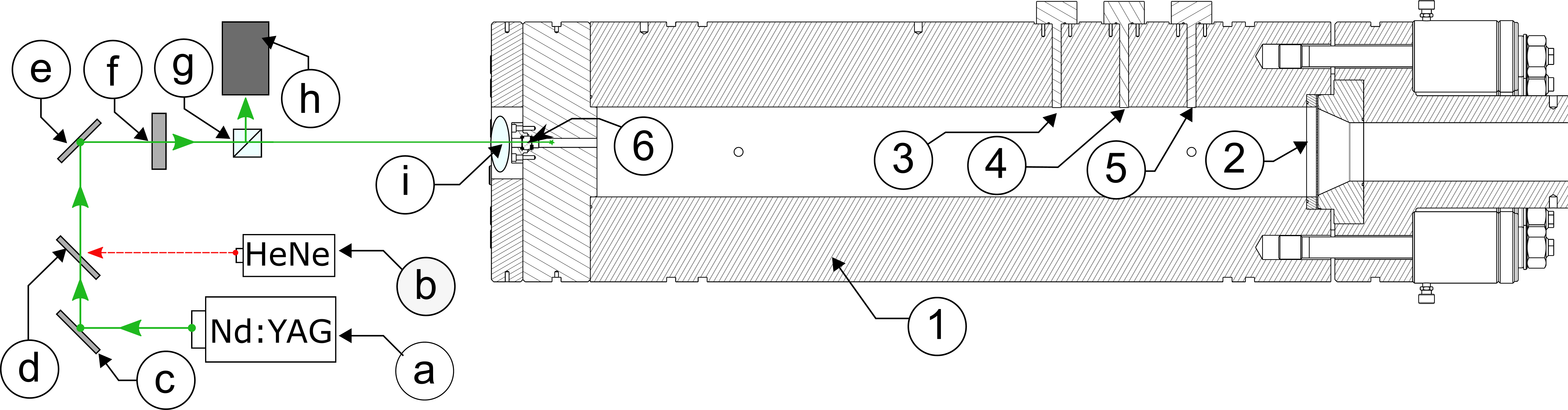}
\caption{ESTHER Combustion Driver Setup. 1-Combustion Chamber; 2-Diapragham/Blank; 3-Gas input port; 4-Gas output port; 5-Kistler gauge; 6-Sapphire Optical Window. a-High-power Nd:YAG laser 1064 nm 5 ns 200 mJ; b-Red diode laser (for alignment); c,d,e-45º mirror; f-Half-wave plate; g-Beam splitter cube; h-Beam dump; i-Bi-convex focusing lens 100 mm}
\label{Fig: ESTHER_Setup}
\end{figure*}

The ESTHER combustion chamber driver operates using a premixed mixture of He:H$_2$:O$_2$ at filling pressures up to 100 bar. Alternatively, He could be replaced by N$_2$. Fig. \ref{Fig: ESTHER_Setup} depicts the experimental setup. The combustion chamber is a cylinder with a total length of 1600 mm, a inner diameter of 200 mm and a wall thickness of 200 mm. Ignition occurs inside a channel with a length of 144 mm and an inner diameter of 20 mm, using a remote-controlled high power Nd:YAG laser and an optical setup of mirrors and lenses. The system is divided into 3 parts, the combustion chamber and its associated optics (1-6), the laser heads (a and b), and the beam conditioning system (c-i). The red and Nd:YAG were first aligned, so that the latter could be  turned off as a safety measure and the alignment made with the former one. The beam from the high power Nd:YAG laser (Quantel Brilliant, 1064 nm, 180 mJ, 5 ns) is reflected by two 45º mirrors (High power CVI mirrors) to adjust its height and azimuthal deviation, passes through a half-wave plate and a beam splitter cube and focusing lens, then enters the chamber. The laser beam power sent to the chamber can be adjusted using the half-wave plate and the polarizer cube.

The combustion is characterized by its pressure signal, measured by a piezoelectric pressure transducer (Kistler 6215) connected to a Charge Amplifier (Kistler 5015). The piezoelectric is positioned into the third port of the driver and protected with a heat shield system (Kistler 6567) as recommended by the manufacturer, in fig.~\ref{Fig: ESTHER_Setup}. The signal is recorded into a digital storage oscilloscope (Tektronics MDO4104B-3) and a FPGA based acquisition board RedPitaya at 125 MSamples/s and variable decimation rate. 

The gas filling system was designed to fill the chamber with an arbitrary mixture of He:H$_2$:O$_2$ up to 100 bar of pressure, identified by number 5 in fig.~\ref{Fig: ESTHER_Setup}. It uses standard 200 bar gas bottles. The option to use N$_2$ instead of He was added to the system. The gas filling system is controlled and monitored by an industrial standard Siemens Programmable Logic Controller (PLC) system (S7-1200 CPU Module 1215C), associated to an EPICS/CS-Studio higher-level control and monitoring layer, operated remotely from a separate control room. The combustion chamber filling pressure is monitored by a pressure transducer is an Ashcroft A4-S-A-F09-42-D0-2000psi-G-XK8-X6B series A4 model, capable of measuring absolute pressures from atmospheric up to 130 bar.

\section{Results and Discussion}

\subsection{Phenomenological description}


A total of 93 shots were performed, 63 with helium acting as diluent and 30 with nitrogen as diluent gas. Direct observation with  Schlieren or shadowgraph photography of the combustion process could not be made because of the large pressures and the thick walls of the vessel. Ignition is formed by an highly reactive plasma kernel \cite{Ronney-1994, Bradley-2004, Ma-1998} near the focal point of the lens (around 40 mm from the sapphire window in our experiment).

\begin{figure}[htb]
\centering
\includegraphics[width= 0.52\textwidth]{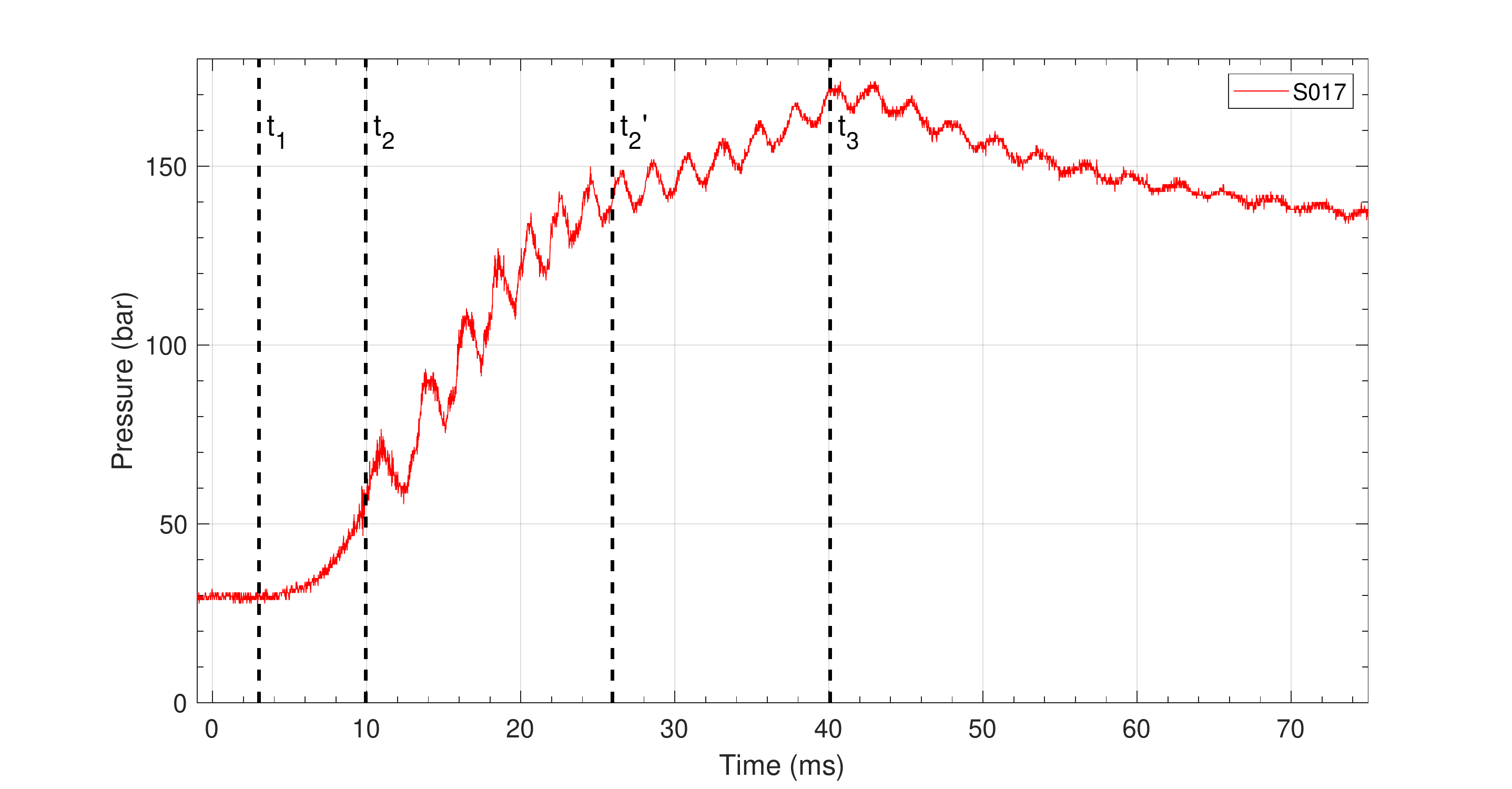}
\caption{S017 pressure signal, $p_0 = 30$ bar, $\lambda=1.2$, $X_{\textrm{He}}=71\%$. Characteristic times $t_2'$ the time when flame front inversion occurs and a transition to a ``tulip flame". }
\label{Fig: S017_Example}
\end{figure}

Fig~\ref{Fig: S017_Example} illustrates the pressure signal of shot S017 with a transition to ``tulip" flame. The laser firing is identified by the flash lamp at $t_0$. Initially, the flame front expands spherically from the ignition channel exit until it reaches the chamber wall, from $t_1$ to $t_2$. As the flame front expands its area, the rate of reaction increases thus accelerating the gas consumption and pressure rise \cite{Liberman-2010}. Then the flame front moves forward at about constant velocity until the gas is totally consumed at $t_3$. An acoustic wave is formed when the flame front hits the chamber outer wall at $t_2$, observed by the oscillation in the pressure signal \cite{Xiao-2014, Xiao-2017}. A transition to a ``tulip" flame can be observed when a second slope is formed at $t_2'$. The values of flame velocity are computed from the pressure signal. The average flame velocity is computed by taking the distance from the exit of ignition channel to the piezoelectric sensor (1346 mm), and the time between the initial rise $t_1$ and peak pressure $t_3$. Faster flames create stronger acoustic waves \cite{Xiao-2014}. 

\begin{figure}[htp]
\centering
\includegraphics[width= 0.5\textwidth]{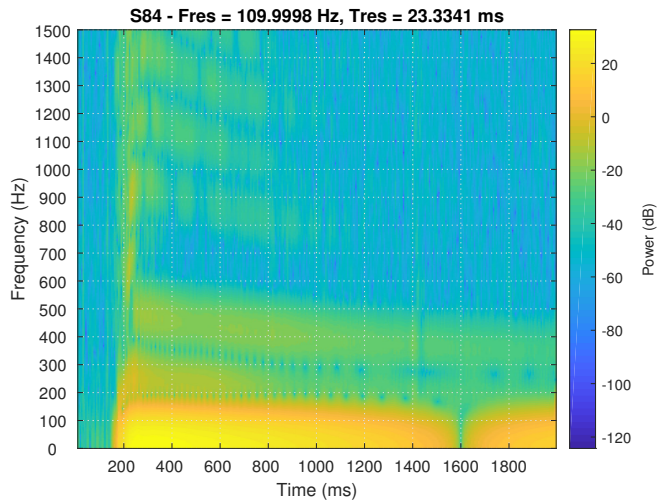}
\caption{Spectrogram (Time-dependent Fast-Fourier transform) of shot S084 (N$_2$:H$_2$:O$_2$, X$_{\textrm{N2}}$=72\%, $\lambda$=1.1, $p_0=10$ bar) pressure signal. A zoom on region up to 1500 Hz is depicted. The region of 150 Hz has a peak corresponding to the acoustic wave pressure oscillation.}
\label{Fig: FFT}
\end{figure}

A spectral analysis of the pressure signal can be done by performing a fast-fourier transform of the pressure signal. A time-dependent Fast-Fourier transform (spectrogram) was computed for shot S084 (N$_2$:H$_2$:O$_2$, X$_{\textrm{N2}}$=72\%, $\lambda$=1.1, $p_0=10$ bar) and depicted in fig.~\ref{Fig: FFT}. Combustion starts slightly before 200 ms, noted by increase in acoustic wave strength. The gas post-combustion will cool down, lowering the sound speed and the acoustic wave frequency. This is observable by the peak frequency shifting to lower values over time. The spectrogram depicts multiple oscillation frequencies, the most crucial are in the 0 to 600 Hz region. The highest frequency decay shortly after being created.

\subsection{Stoichiometry and dilution ratios}

\subsubsection{He/N2 Dilution}

\begin{figure}[htb]
\centering
\includegraphics[width= 0.50\textwidth]{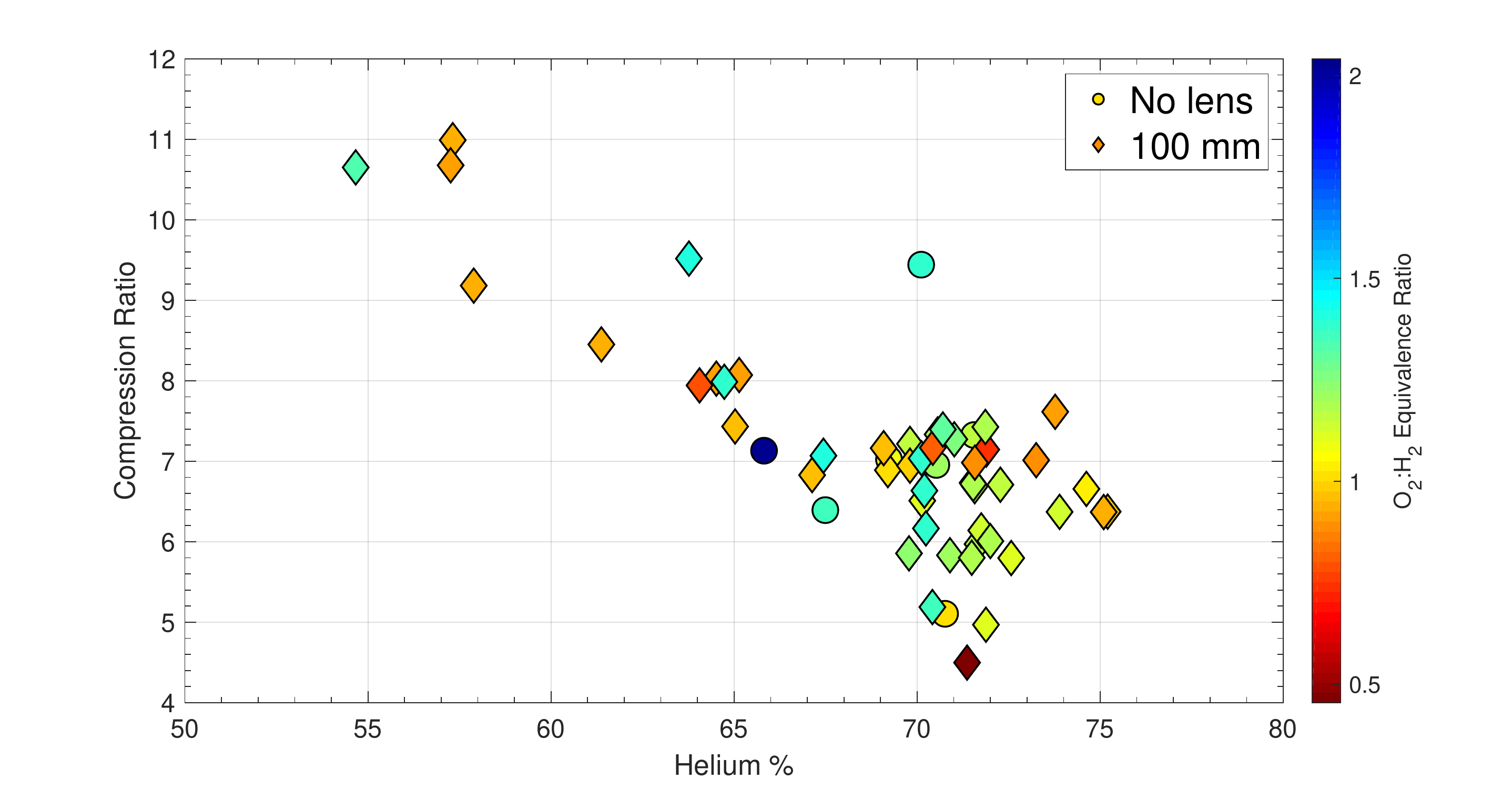}
\caption{Scatter plot of compression function of helium dilution, O$_2$:H$_2$ equivalence ratio in color-axis, focal lens distance in markers symbol.}
\label{Fig: Scatter_CR_He_AirFuel_Lens}
\end{figure}

Four major parameters were analysed during the campaign, the compression ratio/peak pressure, flame velocity, acoustic wave amplitude and combustion mode (deflagration/detonation). Fig.~\ref{Fig: Scatter_CR_He_AirFuel_Lens} illustrates the compression ratio against helium dilution and O$_2$:H$_2$ equivalence ratio in color scheme. As expected, larger helium dilutions decrease the compression ratio. Helium is an inert species which do not take part into the exothermic reaction, nonetheless absorbs part of the release heat. For cases with dilution below 60\%, detonations can occur, since its compression ratio values are above 8.

\begin{figure}[htb]
\centering
\includegraphics[width= 0.50\textwidth]{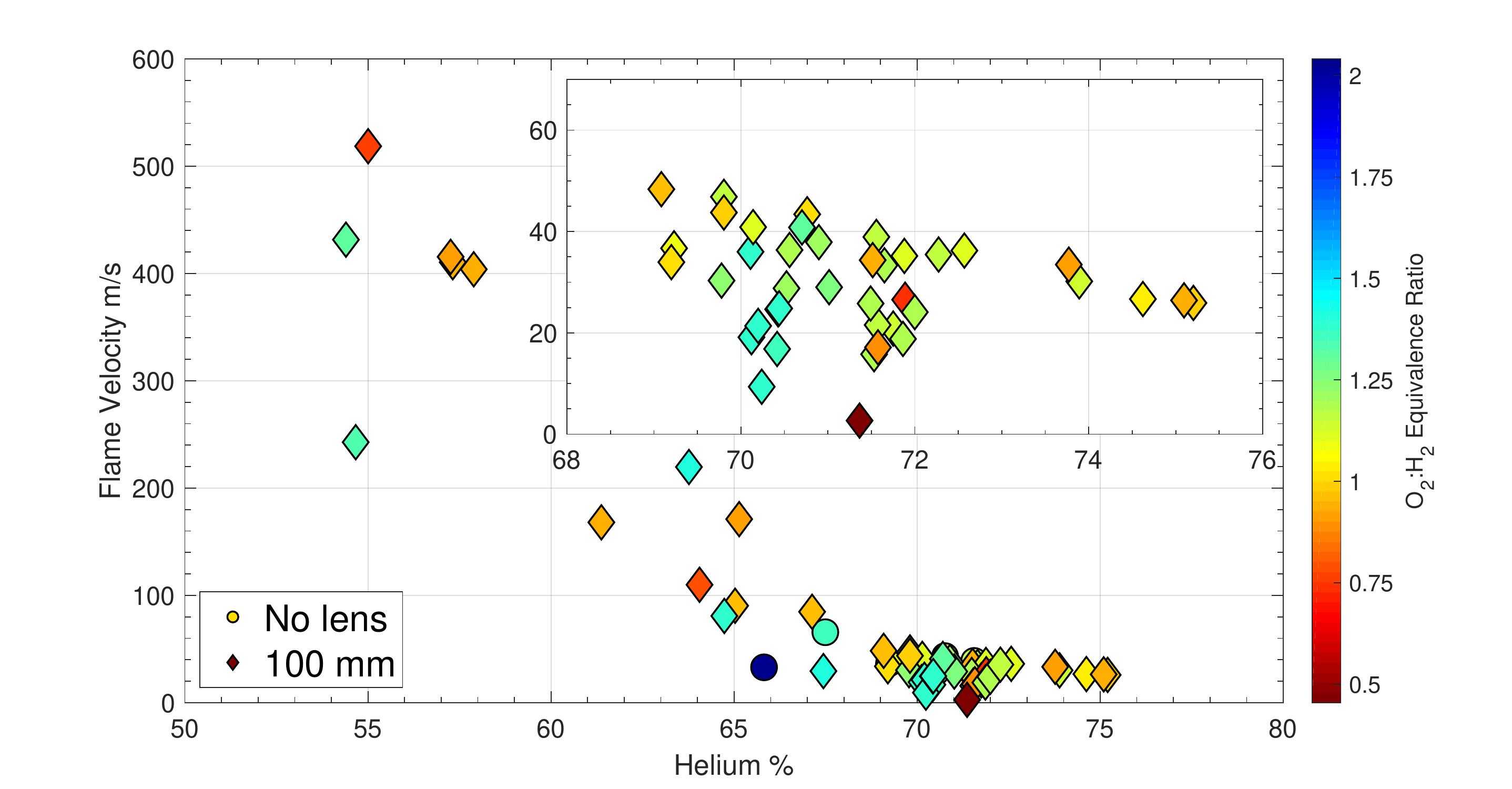}
\caption{Scatter plot of average flame speed function of helium dilution, O$_2$:H$_2$ equivalence ratio in color-axis, focal lens distance in markers symbol.}
\label{Fig: Scatter_FSpeed_He_AirFuel_Lens}
\end{figure}

\begin{figure}[htb]
\centering
\includegraphics[width= 0.5\textwidth]{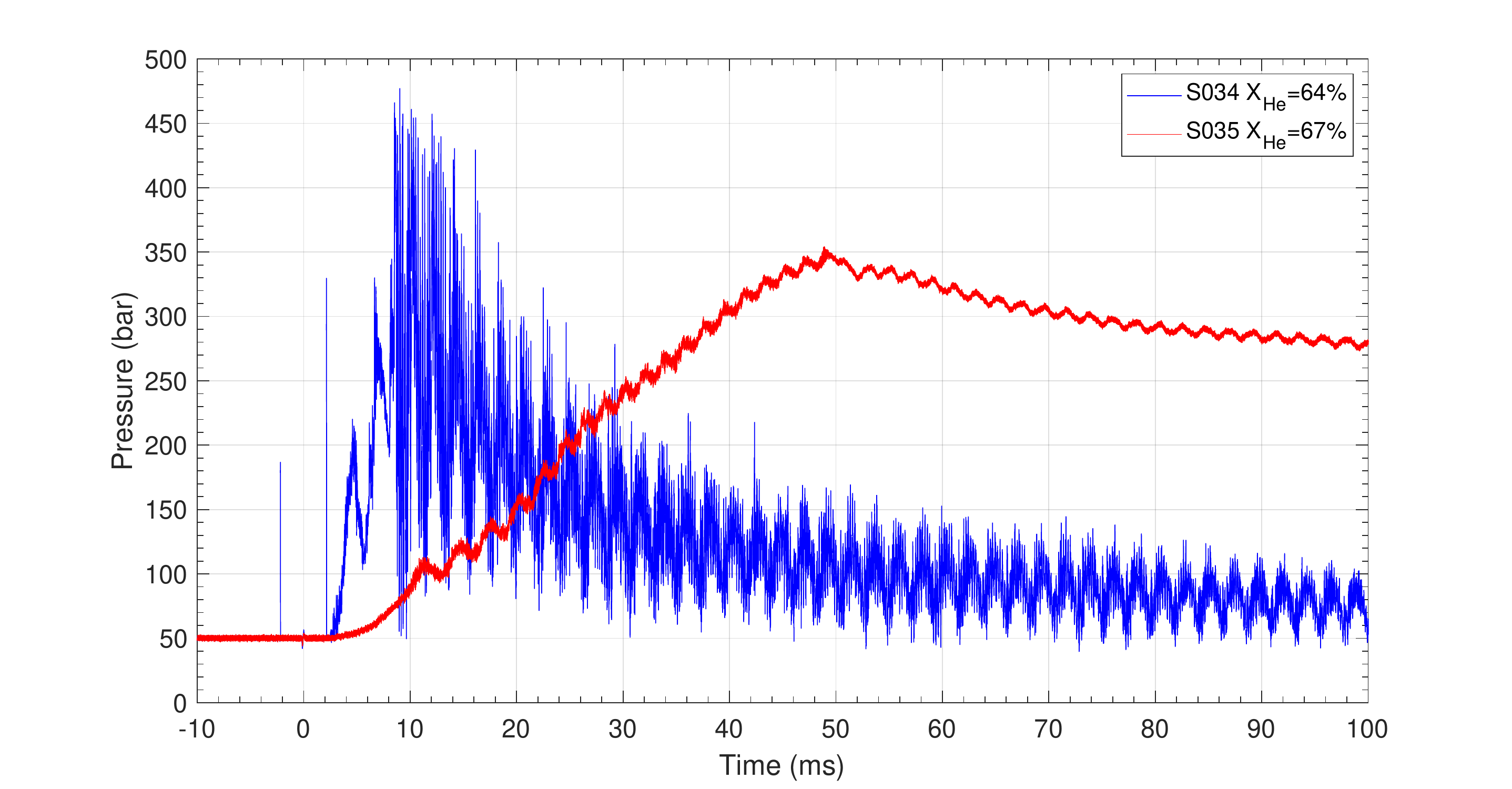}
\caption{Pressure signals for different diluent molar fractions, S034 64\% (blue) and S035 67\% (red). $p_0=$50 bar, $\lambda$=1.40 and lens focal distance 100 mm.}
\label{Fig: Compare_He_S034_S035}
\end{figure}

Fig~\ref{Fig: Scatter_FSpeed_He_AirFuel_Lens} depicts the average flame velocity function of the helium dilution and O$_2$:H$_2$ equivalence ratio. Similarly to the compression ratio, the overall lower temperature will lower the flame velocity. The works of \cite{Tang-2009} agree with our results, where higher dilutions by inert gas lowers the compression ratio and flame velocity. Shots with helium dilutions below 65\% may be yielding detonations, as their average flame velocity is in excess of 200 m/s, the sound speed of the gas inside the chamber. A comparison of shots S034 and S035, $p_0=50$ bar, $\lambda=1.40$ and helium dilutions of 64 and 67\% respectively, pressure signal is shown in fig.~\ref{Fig: Compare_He_S034_S035}. The peak pressure and flame velocity of S035 are lower, due to the higher helium percentage. The amplitude of the acoustic wave is also lower for S035. Ref~\cite{Xiao-2014} relate the acoustic wave amplitude to the flame front velocity, the author stated that larger diameter ducts acoustic waves were stronger due to the higher achieved flame velocities. Nonetheless, the acoustic wave is also dependent on filling pressure and O$_2$:H$_2$ equivalence ratio.

\begin{figure}[htb]
\centering
\includegraphics[width= 0.5\textwidth]{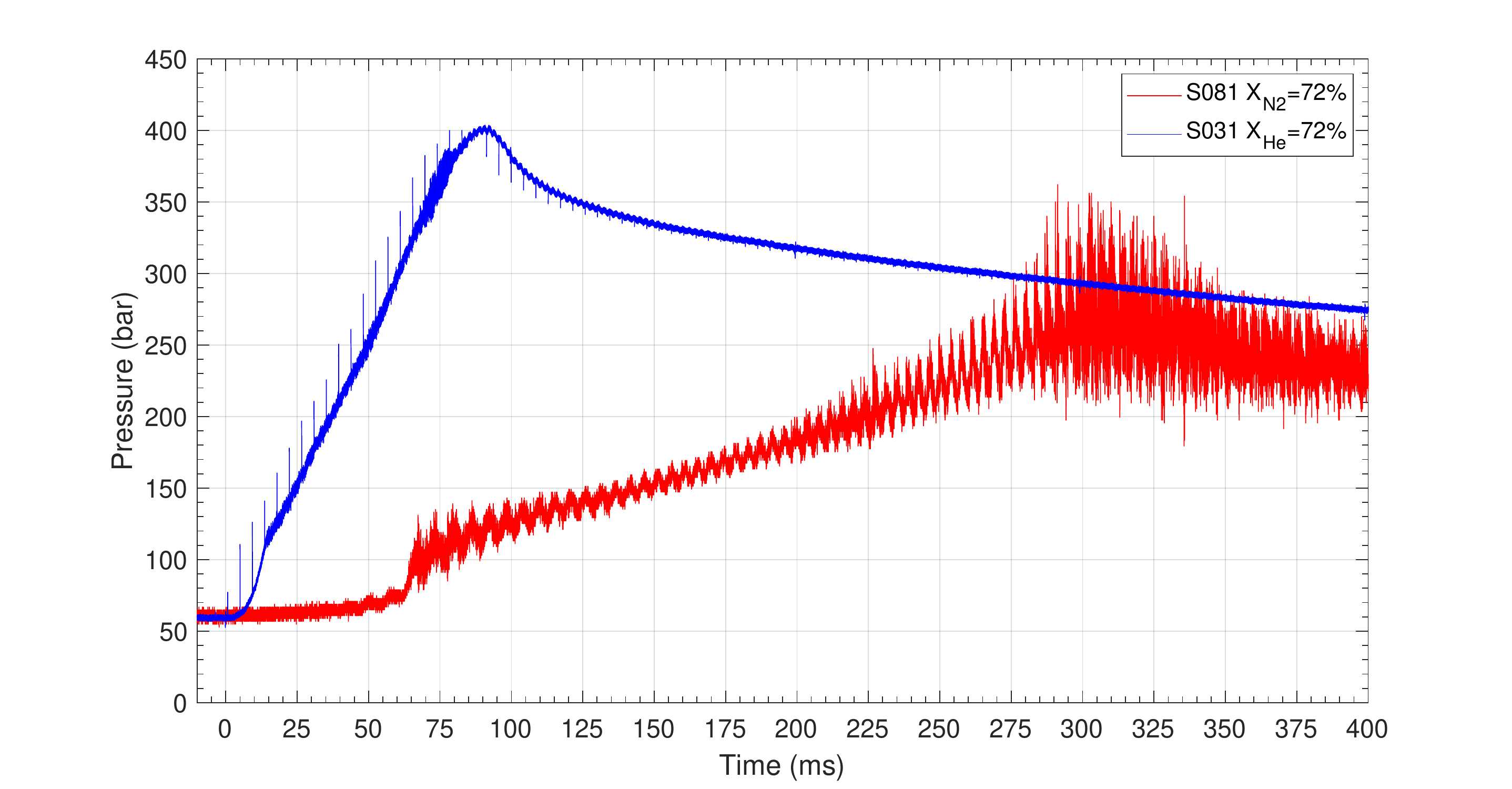}
\caption{Pressure signal comparison for different diluent gases, helium in S031 (blue) and nitrogen in S081 (red). $p_0=$60 bar, $\lambda$=1.20 and lens focal distance 100 mm.}
\label{Fig: He_N2_diluent}
\end{figure}

Nitrogen as a diluent was also tested in place of helium. Both gases have similar behaviour, where increasing the diluent fraction lower compression ratio, flame velocity and acoustic waves amplitude. When comparing helium and nitrogen, the latter is heavier and has a lower specific heat capacity than the former. This means a slower flame velocity and lower compression ratio. Due to the slower flame the wall heat losses increases, further lowering compression ratios when compared to helium diluted shots. The final difference is that helium shots have weak acoustic waves compared to nitrogen ones. Fig~\ref{Fig: He_N2_diluent} compares two shots with different diluent gases S031 (helium) and S081 (nitrogen). Combustion on S081 lasts around 4 times longer than S031. The time lag between the laser pulse and pressure rise is also greater for nitrogen shots. Acoustic waves for nitrogen diluted shots have a higher amplitude and a lower frequency, a consequence of higher mass when compared to helium ones. At pressure peak, strong acoustic waves appear. Because of these undesired strong oscillations the nominal operational mixture ratios had to be shifted from 8:2:1.2 [He:H$_2$:O$_2$] to 10:2:1.2 [N$_2$:H$_2$:O$_2$] in nitrogen diluted shots. These results are in accord to \cite{Burke-2010}, where the influence of different diluents gases on the gas mass burning rate were evaluated.

\subsubsection{O2:H2 Equivalence Ratio}

\begin{figure}[htb]
\centering
\includegraphics[width= 0.45\textwidth]{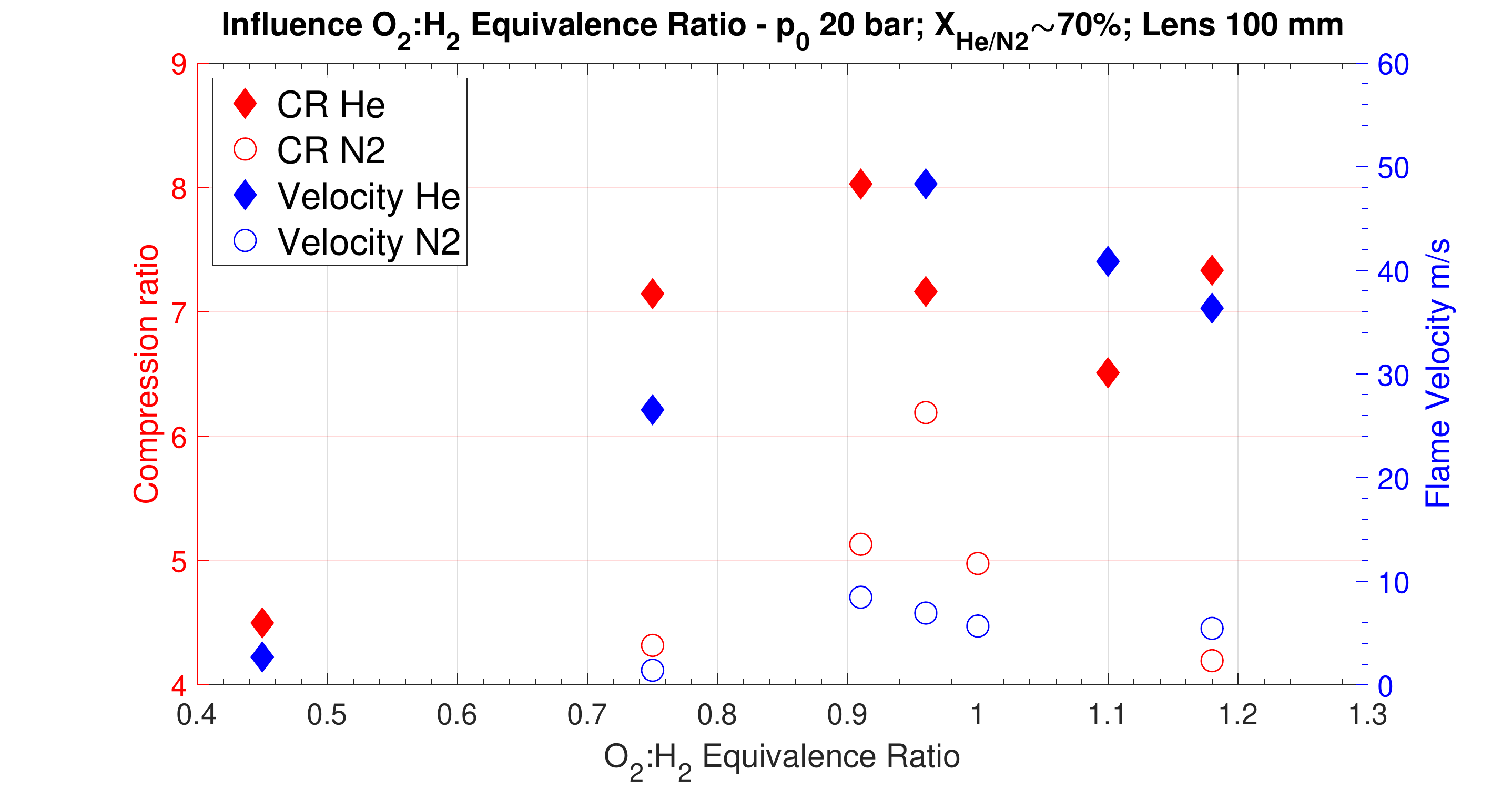}
\caption{Compression ratio function of O$_2$:H$_2$ equivalence ratio (left axis), average flame velocity (right axis). $p_0=$20 bar, X$_{He}$=71\%, lens focal distance 100 mm.}
\label{Fig: CR_AirFuel}
\end{figure}

Figure \ref{Fig: CR_AirFuel} depicts the compression ratio and flame velocity values function of O$_2$:H$_2$ equivalence ratio. Stoichiometric and slightly rich mixtures show higher compression ratio values. Slightly-rich mixtures have higher compression ratios than stoichiometric ones, which might be explained by faster flames which decrease heat losses at the wall. The higher velocity may further be explained by the higher diffusivity of H$_2$, as the works of \cite{Tang-2009, Burke-2007, Burke-2010, Santner-2013, Prasad-2017} report. Nonetheless, if the mixture goes too rich then the heat release will decay, thus will the temperature, compression ratio and flame velocity.

\begin{figure}[htb]
\centering
\includegraphics[width=0.5\textwidth]{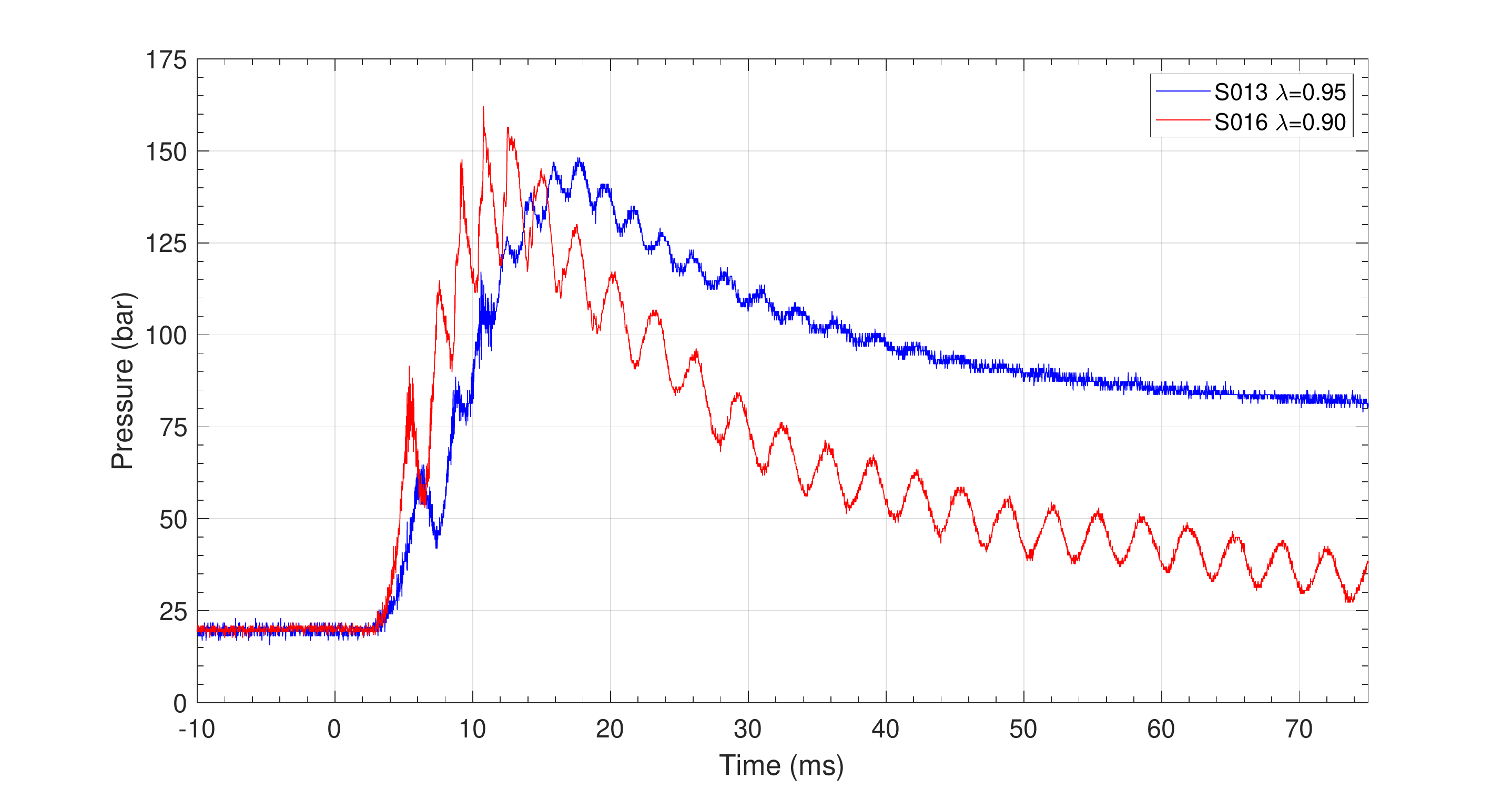}
\caption{Influence of O$_2$:H$_2$ equivalence ratio on pressure signal comparison of shots at $p_0$=20 bar, X$_{\textrm{He}}$ of 65\% }
\label{Fig: Compare_20bar_AirFuel}
\end{figure}

Fig.~\ref{Fig: Compare_20bar_AirFuel} shows the pressure signal of shots S013 and S016, $p_0=20$~bar, X$_{\textrm{He}}$=65\%, $\lambda$=0.95 and 0.90, respectively. The peak pressure is similar in  both cases, however, S016 has a faster combustion than S013. Thus, S016 creates stronger acoustic waves, in concordance with \cite{Xiao-2017}.

\begin{figure}[htb]
\centering
\includegraphics[width=0.5\textwidth]{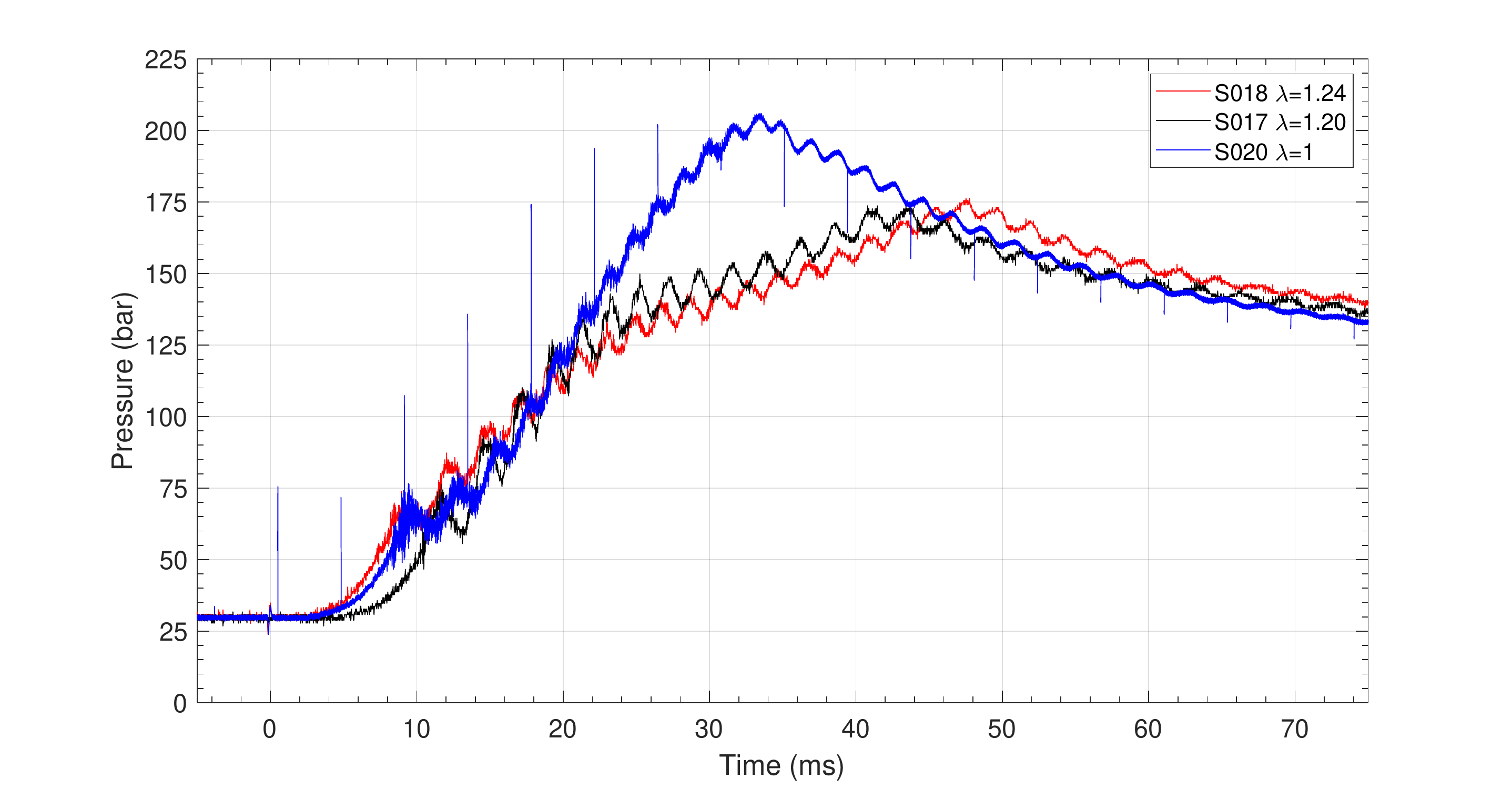}
\caption{Pressure signal comparison for three shots $p_0$=30 bar, X$_{\textrm{He}}\simeq$71\% and lens focal distance of 100 mm. For lean mixtures ($\lambda>$1) a double slope signal is observed, meaning a transition to ``tulip" flame occurred.}
\label{Fig: Compare_30bar_AirFuel}
\end{figure}

\begin{figure}[htb]
\centering
\includegraphics[width= 0.5\textwidth]{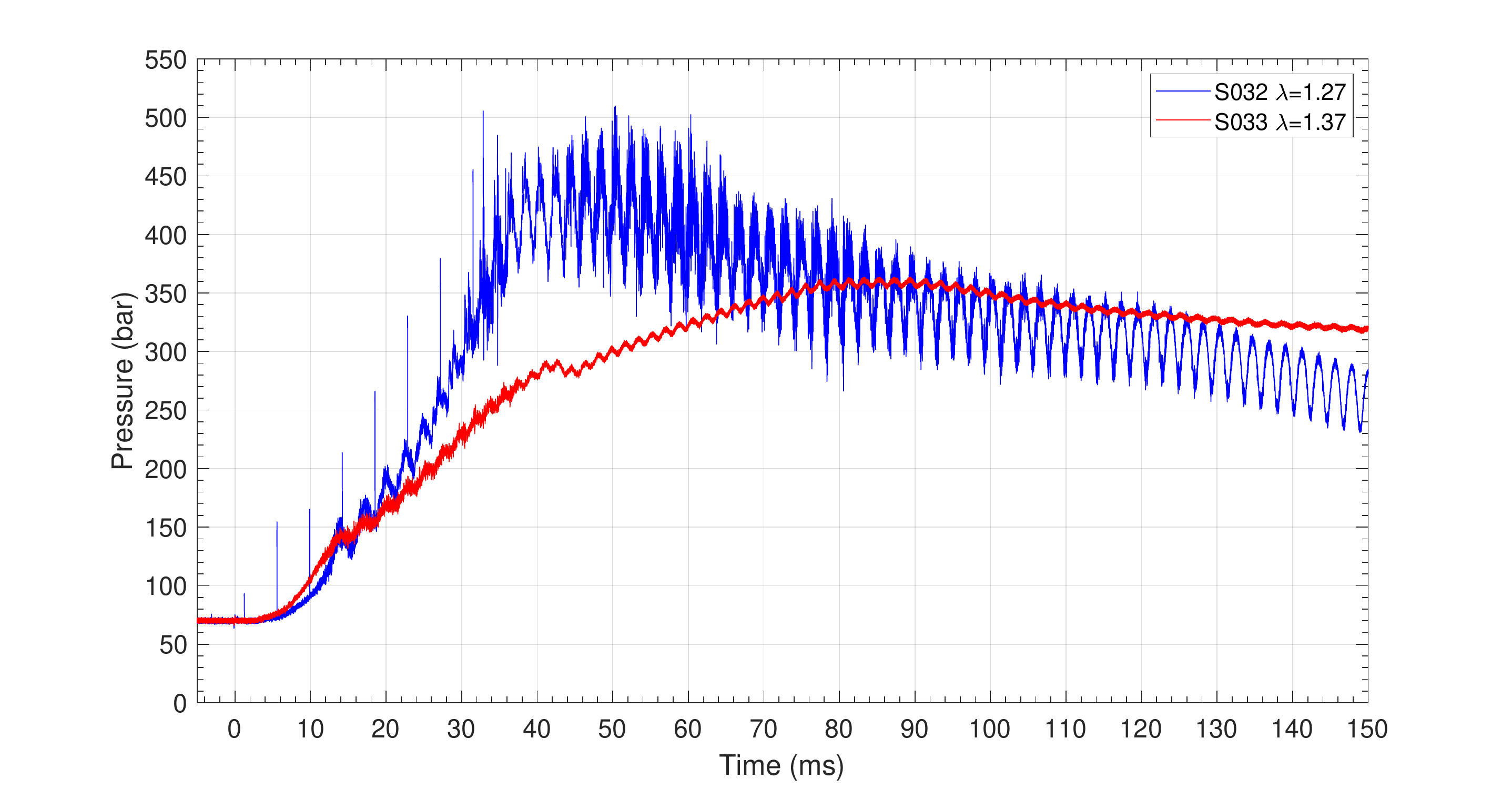}
\caption{Influence of O$_2$:H$_2$ equivalence ratio on pressure signal comparison of shots at $p_0$=70 bar, X$_{\textrm{He}}=$71\%.}
\label{Fig: Compare_70bar_AirFuel}
\end{figure}

A comparison of three 30 bar shots with helium dilution of 70\% is depicted in fig.~\ref{Fig: Compare_30bar_AirFuel}. Both lean mixtures have a double slope and take lower to complete combustion. Lastly, fig.~\ref{Fig: Compare_70bar_AirFuel} compares two lean mixture, $\lambda$=1.27 and 1.37, shots at 70 bar and 71\% helium dilution. The mixture closer to stoichiometric has higher peak pressure and faster combustion, as well, as a much stronger acoustic wave. All the results agree with the expected U-curve profile for both compression ratio and flame velocity.

\begin{figure}[htb]
\centering
\includegraphics[width=0.5\textwidth]{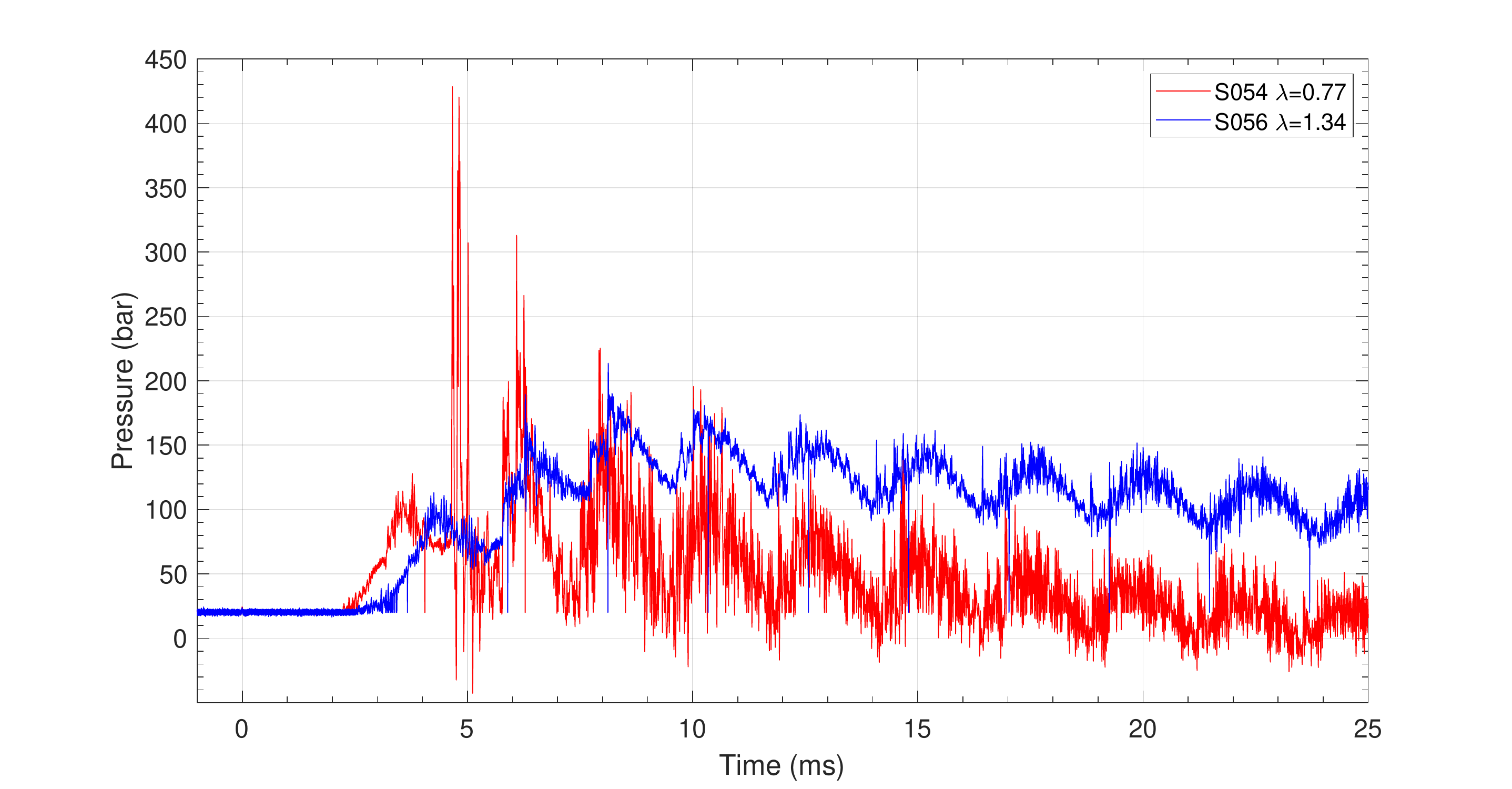}
\caption{Pressure signal comparison for shots at $p_0=$20 bar, X$_{\textrm{He}}=$55\% and lens focal distance of 100 mm. A detonation is observable for S054 ($\lambda$=0.77) but not for S056 ($\lambda$=1.34).}
\label{Fig: AirFuel_Detonation}
\end{figure}

Ref~\cite{Lee-1984} reports that transition to detonation are easier in stoichiometric conditions. In the case of hydrogen mixtures, rich and stoichiometric mixtures detonate or transit to a detonation easier than lean ones. Fig.~\ref{Fig: AirFuel_Detonation} illustrates this phenomena, where the rich $\lambda$=0.77 shot transits to a detonation and the lean $\lambda$=1.34 does not.

\subsection{Pressure effects}

\begin{figure}[htb]
\centering
\includegraphics[width= 0.5\textwidth]{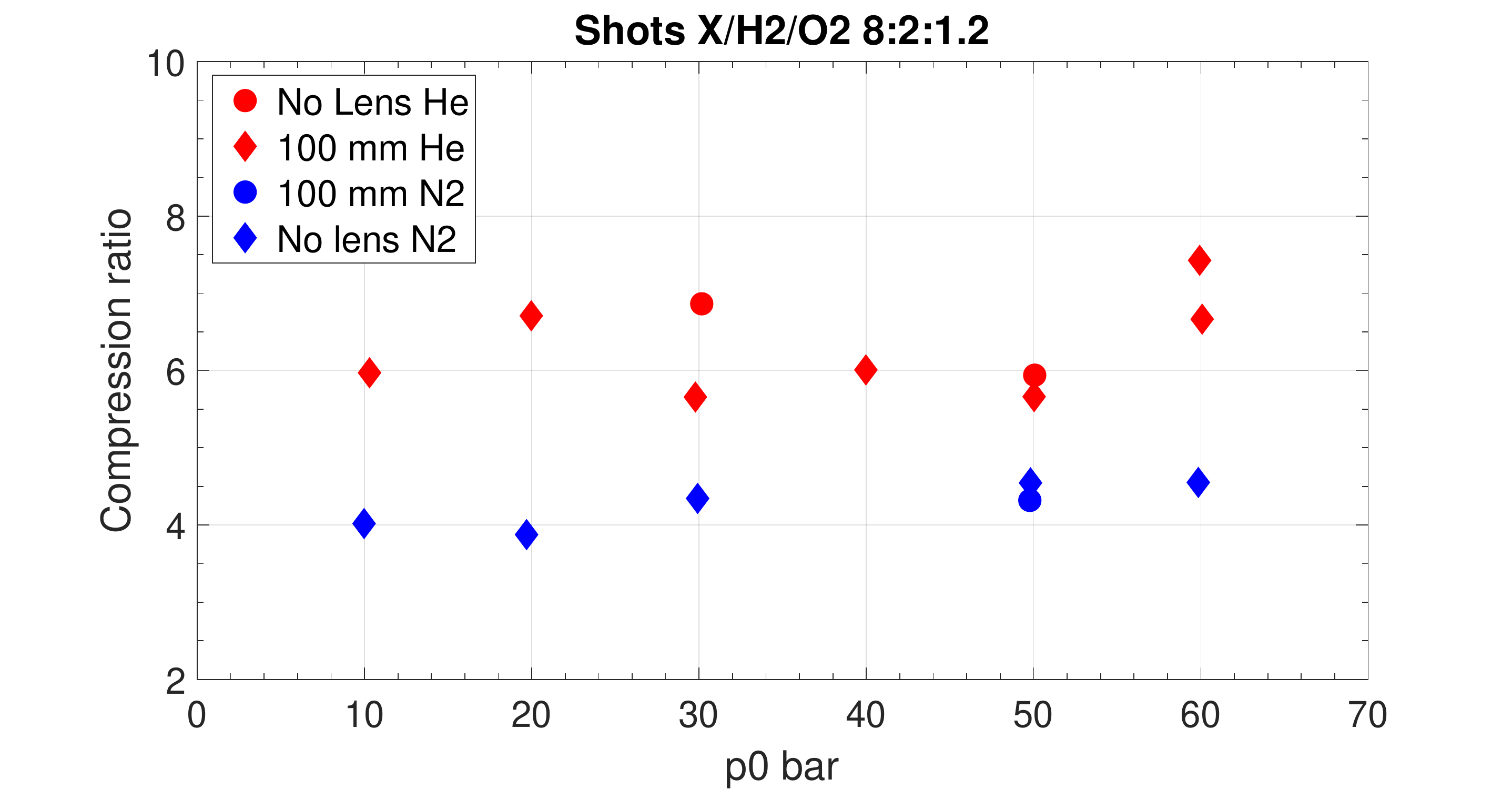}
\caption{Compression ratio comparison for the ``typical" [X:H$_2$:O$_2$] 8:2:1.2 shot at different filling pressures.}
\label{Fig: Compare_8_2_1.2_CR}
\end{figure}

\begin{figure}[htb]
\centering
\includegraphics[width= 0.5\textwidth]{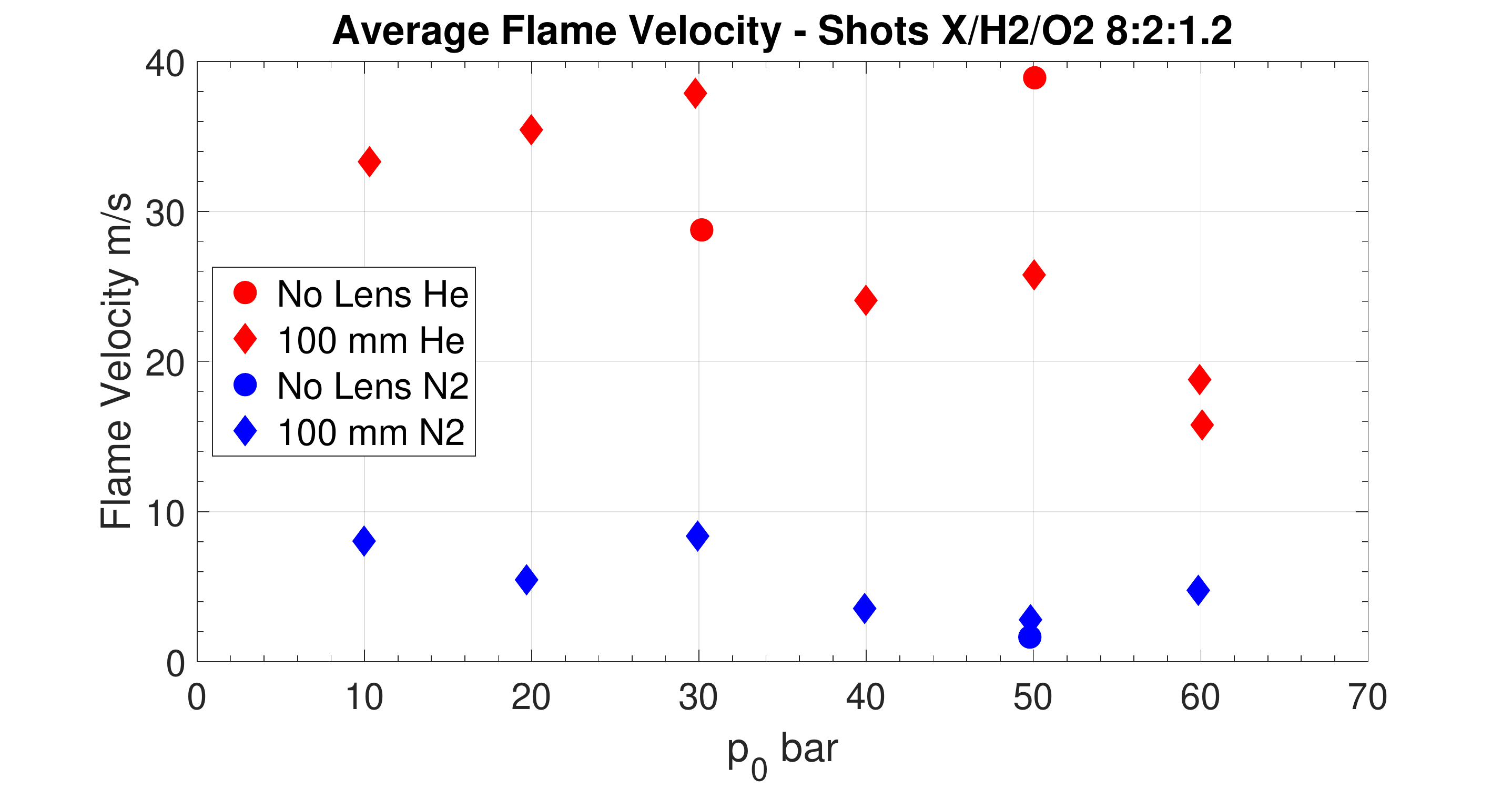}
\caption{Average flame velocity comparison for the ``typical" [He:H2:O2] 8:2:1.2 shot at different filling pressures.}
\label{Fig: Compare_8_2_1.2_Vel}
\end{figure}

Fig.~\ref{Fig: Compare_8_2_1.2_CR} depicts the compression ratio values for helium and nitrogen diluted mixtures at different filling pressures. Compression ratio does not significantly change with the pressure, as in the works of \cite{Tang-2009}. The difference in compression ratio might be explained by the weaker acoustic wave when the 100 mm focal lens is employed. The effects of focal lens are discussed in detail further down. Fig.~\ref{Fig: Compare_8_2_1.2_Vel} shows the velocity values for the shots in fig.~\ref{Fig: Compare_8_2_1.2_CR}.

The slower flames at higher pressures are in agreement with \cite{Burke-2007, Burke-2010, Kuznetsov-2011, Kuznetsov-2012, Tse-2001, Prasad-2017, Santner-2013}. Due to the high dilution factor in our experiment $\sim$71\%, it was anticipated that higher pressures would lower the flame velocity. References~\cite{Burke-2010, Kuznetsov-2011} explain that by the change in kinetic scheme, from a chain-branching to a straight-chain one. 

\begin{figure}[htb]
\centering
\includegraphics[width= 0.5\textwidth]{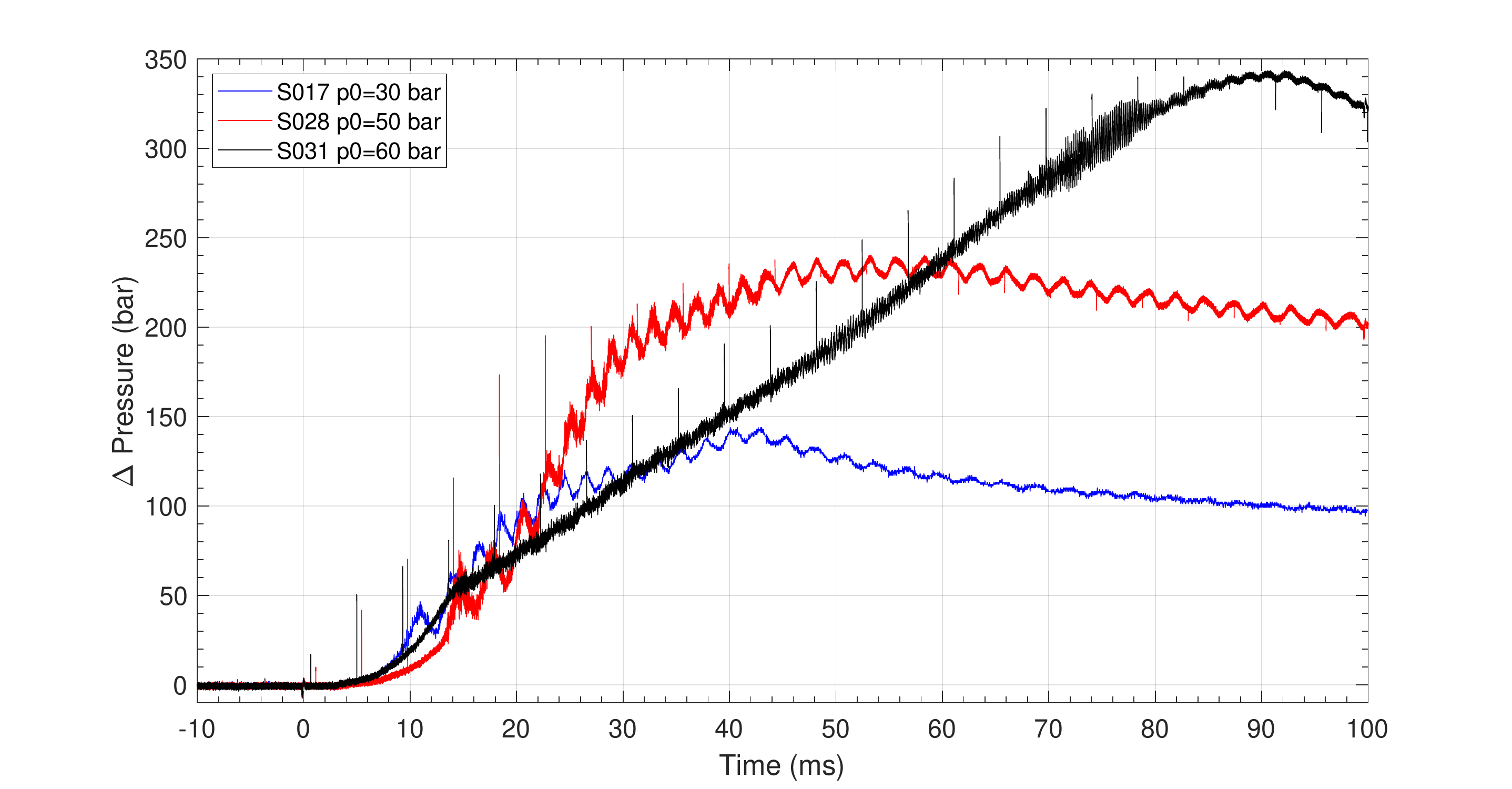}
\caption{Comparison [He:H$_2$:O$_2$] 8:2:1.2 shots at different filling pressures}
\label{Fig: Compare_p0_S017_S028_S0231}
\end{figure}

The pressure signal comparison for shots S017, S019 and S031 with a 8:2:1.2 [He:H$_2$:O$_2$] can be seen in fig.~\ref{Fig: Compare_p0_S017_S028_S0231}. [He:H$_2$:O$_2$] mixture at different pressures. The lower pressure shots, S017 and S028, show a``double" slope combustion, with the second slope starting at around t=25 ms and t=30 ms, respectively. The highest filling pressure shot S031 $p_0=60$ bar however had a single slope combustion behaviour. 

\begin{figure}[htb]
\centering
\includegraphics[width=0.5\textwidth]{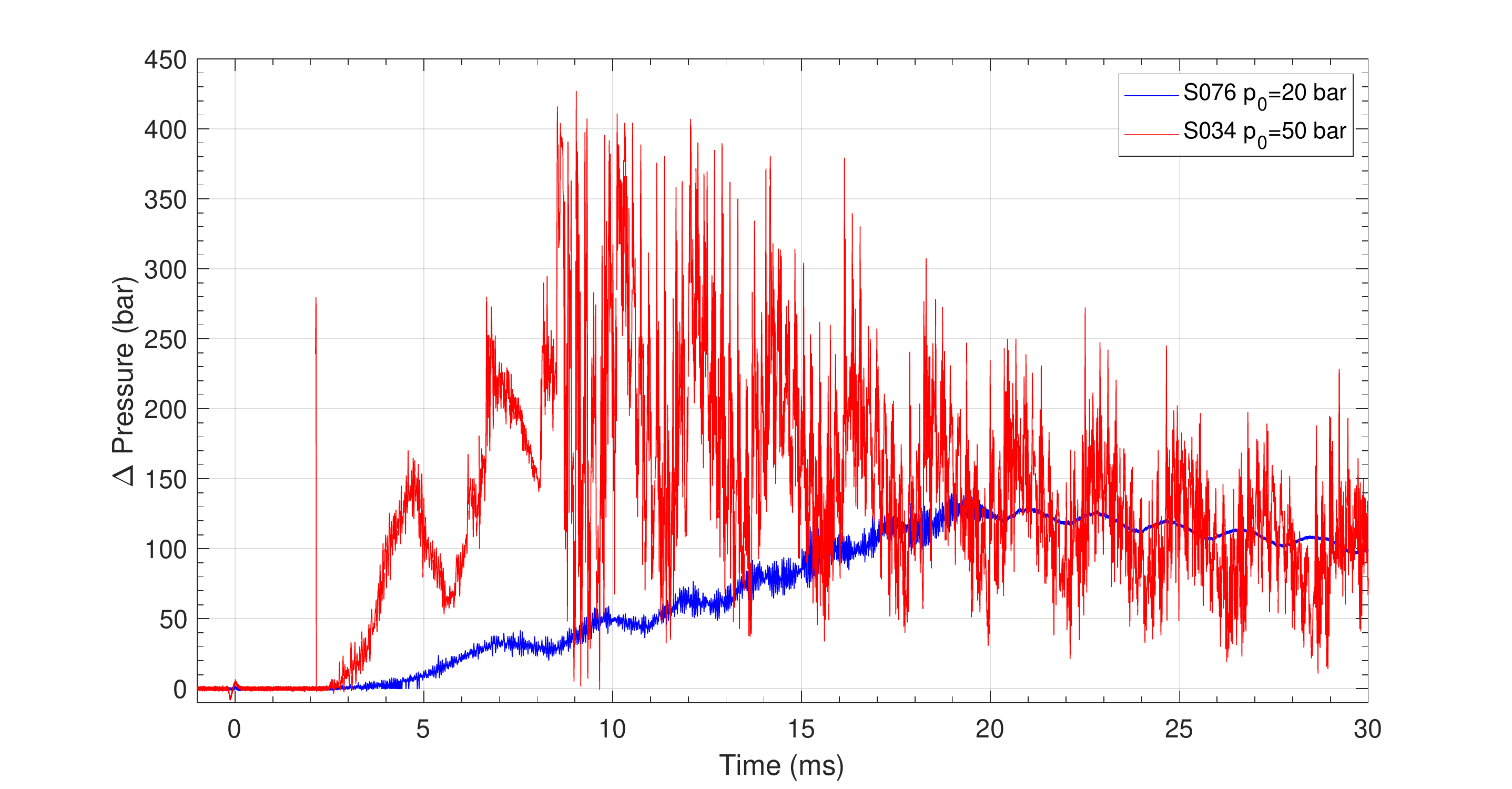}
\caption{Pressure signal comparison for S034 ($p_0$=50 bar) and S076 ($p_0$=20 bar), $\lambda$=1.40, $X_{\textrm{He}}$=64\%.}
\label{Fig: Deton_p0_S034_S076}
\end{figure}

\begin{figure}[htb]
\centering
\includegraphics[width= 0.5\textwidth]{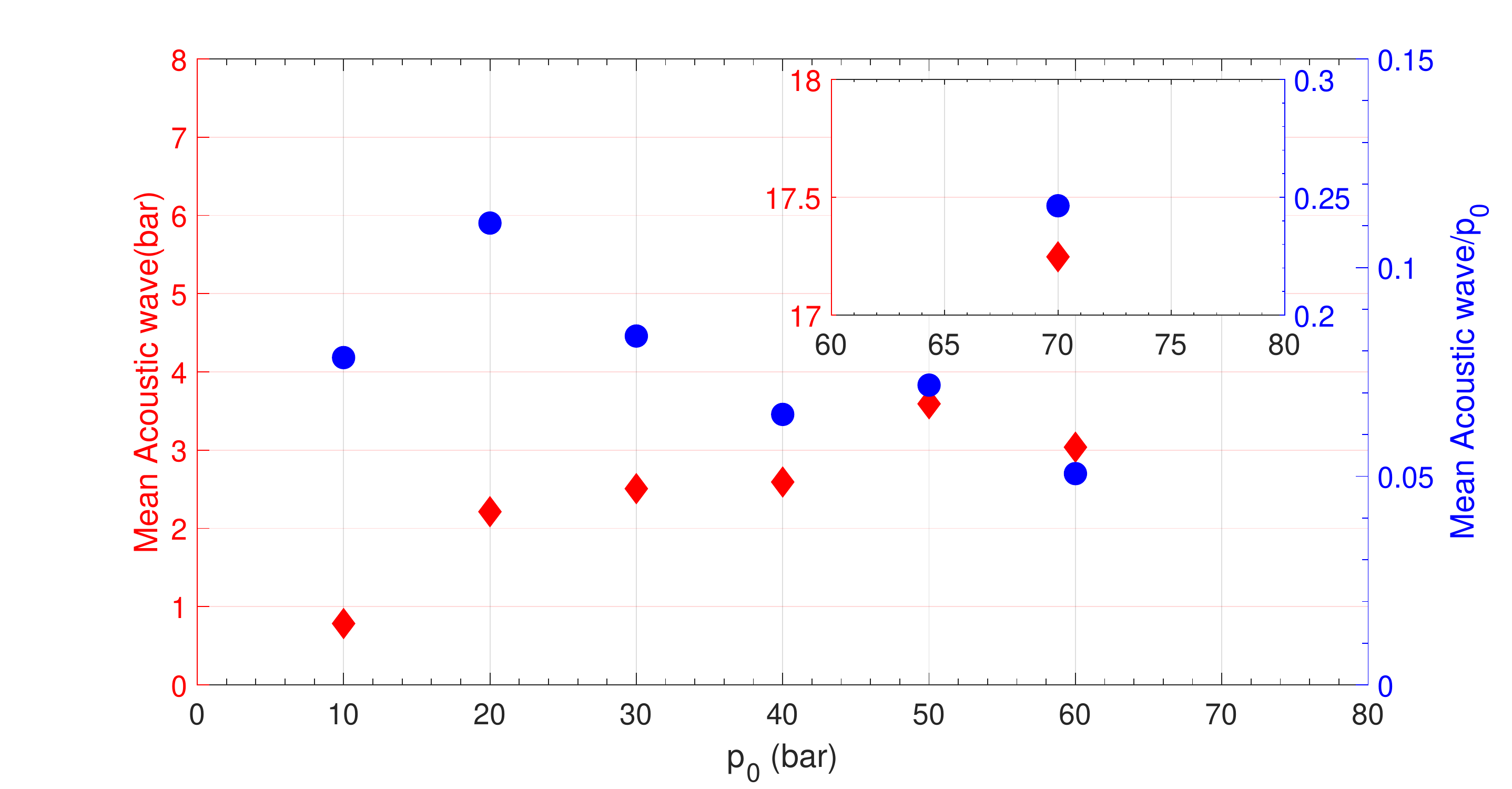}
\caption{Influence of filling pressure on the mean amplitude for acoustic waves. Comparison of 8:2:1.2 [He:H$_2$:O$_2$] mixtures}
\label{Fig: Acoustic_wave_p0}
\end{figure}

Increased pressure would also create stronger acoustic waves and more detonation prone mixtures. This effect is observable in fig.~\ref{Fig: Deton_p0_S034_S076} which compares two shots with similar chemical composition ($\lambda$=1.40, $X_{\textrm{He}}$=64\%) at different pressures. The higher pressure signal transits to a detonation at t$\sim$8 ms. The higher filling pressure decrease the size of detonation cells, thus making the mixture more detonation prone \cite{Knystautas-1982}. The mean acoustic wave amplitude values for different shots is depicted in fig.~\ref{Fig: Acoustic_wave_p0}. Higher filling pressures have increased acoustic waves, a steep rise is observable from 60 to 70 bar filling pressure.

\subsection{Influence of laser focusing}


The laser ignition system operate in two modes: focused beam via a 100 mm bi-convex focusing lens and unfocused beam by removing the aforementioned lens. Using the 100 mm lens the ignition point would occur inside the ignition channel. Without the lens the ignition point of origin could not be exactly determined. For filling pressures below 30 bar ignition required the lens to be mounted; in the range 30 to 60 bar ignition without lens could occur, yet could take multiple laser pulse to initiate; above 60 bar a single pulse would lead to ignition. 

\begin{figure}[htb]
\centering
\includegraphics[width=0.5\textwidth]{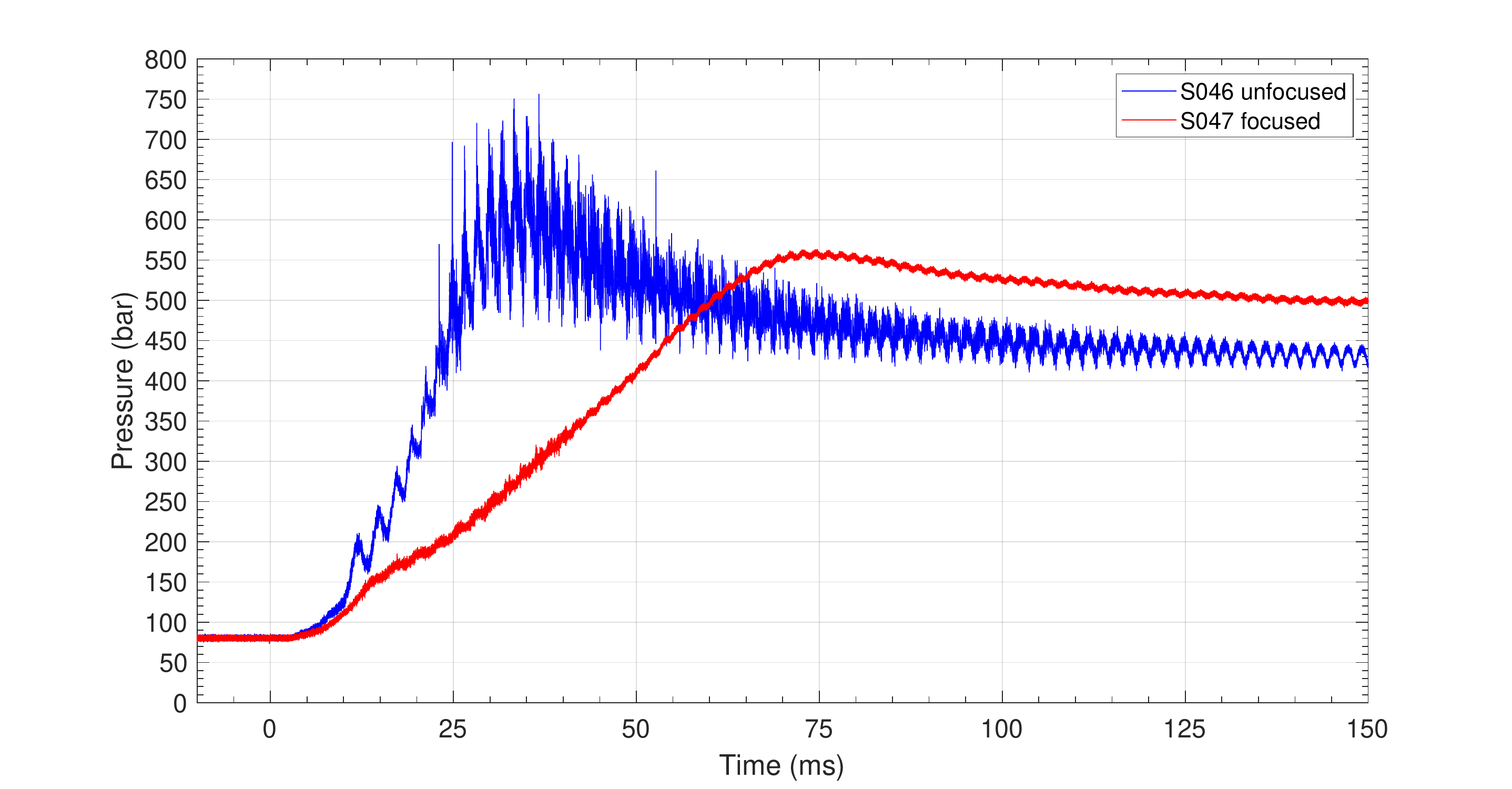}
\caption{Shots S047 and S046 pressure signal comparison, $p_0=80$ bar, $\lambda=1.39$ and X$_{\textrm{He}}=70\%$}
\label{Fig: Compare_Focus_80bar}
\end{figure}

A comparison of the effects of focusing lens is shown in fig.~\ref{Fig: Compare_Focus_80bar}. It compares pressure signal of shots S046 and S047, $p_0=80$ bar, $\lambda=1.39$ and X$_{\textrm{He}}=70\%$, unfocused and focused, respectively. The focused ignition reduces both the acoustic oscillation and the peak pressure values, as well as the average flame velocity. An hypothesis for this is that unfocused ignition may create multiple ignition points along the beam path. From there, multiple flame fronts propagate simultaneously, which creates more instabilities therefore accelerating the gas burning rate. Ignition might also be taking place along a line, thus reducing the characteristic burn length from the chamber length to its diameter. In any case, the faster combustion reduces wall heat losses, thus increasing the peak temperature/pressure of the gas. As stated before, faster flames are also responsible for stronger acoustic oscillations, as observable in fig.~\ref{Fig: Compare_Focus_80bar}. Further research is required to understand these differences, and the inclusion of fast imaging diagnostics appears unavoidable for it. The explanation for how unfocused ignition is achieved for high-pressure combustible mixtures is yet not fully understood. References~\cite{Weinrotter-2004,Phuoc-2006,Kopecek-Reider-2003,RFerreira-Laser-Ignition-ArXiv} show that higher filling pressures lower the minimum pulse energy. The electron cascade most likely starts with ionization of microparticles impurities such as dust or soot. 


\section{Final Remarks}

We have performed experiments on high-pressure He:H$_2$:O$_2$ and N$_2$:H$_2$:O$_2$ evaluating the compression ratio, acoustic wave formation, flame velocity and transition to detonation. The parameters in study were filling pressure [10-100] bar, oxygen-hydrogen equivalence ratio [0.45-2.04], inert gas dilutions of [54-79]\% and laser ignition mounted/unmounted 100 mm bi-convex lens. The effects can be summarized as:

Increased dilution:
\begin{itemize}
\item Lower compression ratio
\item Slower combustion
\item Weaker acoustic waves
\item Less detonation prone mixtures
\end{itemize}

When replacing helium for nitrogen as dilutant gas, the compression ratio and flame velocity further decrease. However, the acoustic waves amplitude is larger. 

Mixtures with O$_2$:H$_2$ equivalence ratio closer to the stoichiometric value were faster and with higher peak pressure. Slightly rich mixtures are faster and with higher compression ratios. These are also more prone to detonation and with stronger acoustic waves. 

Increased filling pressure:
\begin{itemize}
\item Similar compression ratio
\item Slower combustion
\item Stronger acoustic waves
\item More detonation prone mixtures
\end{itemize}

The removal of focusing lens creates stronger acoustic waves and faster combustions. However, laser ignition without the lens can happen for filling pressure above 30 bar. 

A direct imaging of the combustion process is important step to acquired information on the flame front shape and transition to tulip flame. Alongside it, to understand how the ignition occurs along the laser beam path in the un-focused mode.

Operationally-speaking, stable, repeatable deflagrations have been reached for filling pressures in the 10-100 bar range, allowing successful operation of the driver up to 650 bar post-shot pressures, with a high degree of cleanliness.

\bibliographystyle{plain} 
\bibliography{references}

\end{document}